\documentclass[12pt,preprint]{aastex}
\shorttitle{HeliumLetter}
\shortauthors{Nataf et al.}
\usepackage{float}
\usepackage{sidecap}

\begin{document}

\title{Reconciling the Galactic Bulge Turnoff Age Discrepancy with Enhanced Helium Enrichment}

\author{David M. Nataf\altaffilmark{1}, Andrew P. Gould\altaffilmark{1}}
\altaffiltext{1}{Department of Astronomy, Ohio State University, 140 W. 18th Ave., Columbus, OH 43210}

\begin{abstract}
We show that the factor $\sim$2 discrepancy between spectroscopic and photometric age determinations of the Galactic bulge main-sequence turnoff can be naturally explained by positing an elevated helium enrichment for the bulge relative to that assumed by standard isochrones. Helium-enhancement relative to standard isochrones is confirmed at the 2.3$\sigma$ level. We obtain an upper bound on the helium enrichment for the metal-rich ([Fe/H]$\approx+0.30$) stars of ${\Delta}Y\approx+0.11$ relative to canonical expectations, given the requirement that the spectroscopic and photometric ages be consistent and the limiting condition of instantaneous star formation. We discuss phenomenological evidence that the bulge may have had a chemical evolution that is distinct from the solar neighborhood in this manner, and we make several testable predictions. Should this emerging picture of the bulge as helium-enhanced hold, it will require the development of new isochrones, new model atmospheres, and modified analysis and cosmological interpretation of the integrated light of other bulges and elliptical galaxies. 
\end{abstract}

\section{Introduction}
\label{sec:Introduction}
The mean age and age-spread of the Galactic bulge, $t_{\rm{Bulge}}$ and ${\Delta}t_{\rm{Bulge}}$, are fundamental parameters of Galactic evolution, with hierarchical formation theories predicting an older bulge with more rapid formation timescale than models where bulge formation was linked to early disk \citep{2004ARA&A..42..603K,2011A&A...525A.126C,2011arXiv1109.2898I}.  However, age determinations of bulge stars are currently in a state of dissonance, with spectroscopic ages extending to much lower ages than photometric determinations. 

A seemingly secure picture had emerged from \textit{Hubble Space Telescope (HST)} photometry of the Galactic bulge main-sequence turnoff (MSTO). \citet{2002AJ....124.2054K} and \citet{2003A&A...399..931Z} both showed that the bulge MSTO is dim relative to the bulge horizontal branch (HB), with a brightness offset indistinguishable from that found in Galactic globular clusters (GCs), suggesting an old age. \citet{2011ApJ...735...37C} refined this approach by searching for eclipsing binary blue-straggler stars near the MSTO. They found that whatever sparse population of bright stars there was could be entirely explained by bulge blue stragglers, and concluded that no more than 3\% of bulge stars are younger than 5 Gyr. Thus, the statement that the bulge stellar population is old appeared robust: bulge MSTO stars are dim. 

However, this picture has been challenged by  \citet{2010A&A...512A..41B,2011A&A...533A.134B}, who used gravitational microlensing events to obtain high-resolution spectra of bulge MSTO and subgiant (SGB) stars, which yielded detailed measurements of $\log{g}$, $T_{\rm{eff}}$, [Fe/H] and 12 additional chemical abundances. Their sample, microlensed source stars (only $\sim$3\% of microlensed sources are expected to be disk stars) with coordinates $(|l| \lesssim 5, 2 \lesssim |b| \lesssim 5)$ and radial velocity dispersion of $\sim$100 km/s, is  expected to be a predominantly bulge sample. Comparison of the measurements to isochrones led to a big surprise: a vast population of young stars. Seven of their 26 spectra (27\%) yield ages, $t_{\rm{Inferred}}\leq5$ Gyr, an outcome with probability $P < 10^{-5}$ if one assumes the 3\% constraint. This method may seem complex at first glance, but upon close inspection it also appears robust. Consider the MSTO star MOA-2008-BLG-311S. At its high metallicity, [Fe/H]$=+0.36\pm0.07$, standard isochrones \citep{2008ApJS..178...89D} predict it should be no hotter than $T_{\rm{eff}} \approx 5450 K$ if one assumes an old age. However, its measured temperature is $T_{\rm{eff}} = 5944 \pm 68 K$, yielding the spectroscopic fit $t_{\rm{Inferred}} \sim 2.9$ Gyr. Next, consider the metal-rich ([Fe/H]$=+0.37\pm0.05$) SGB star MOA-2009-BLG-259S. At its temperature, $T_{\rm{eff}} = 4953 \pm 93 K$, the expected surface gravity is $\log{g} \approx 3.90$. However, the measured value is $\log{g}=3.40 \pm 0.24$, for $t_{\rm{Inferred}} \sim 3.0$ Gyr. Therefore, the paradigm-challenging data reported by \citet{2010A&A...512A..41B,2011A&A...533A.134B} pass the consistency test of being independently demonstrated in two distinct phases of stellar evolution. In summary:
\begin{itemize}
 \item \textit{HST} photometry: a faint MSTO, indicating an old stellar population \citep{2002AJ....124.2054K,2003A&A...399..931Z,2010ApJ...725L..19B,2011ApJ...735...37C}. The similarity with Galactic GCs suggests $t_{\rm{Bulge}} \approx$ 12.8 Gyr \citep{2009ApJ...694.1498M}. 
 \item High-resolution spectroscopy of [Fe/H]$>$0 stars: high  $T_{\rm{eff}}$ for the MSTO and low $\log{g}$ for the SGB, indicating a young stellar population \citep{2010A&A...512A..41B,2011A&A...533A.134B}. They find a weighted mean age of 5.9 Gyrs for the super-solar metallicity stars.
\end{itemize}
In this \textit{Letter} we suggest that the origin of this discrepancy is with the isochrones that are used to interpret the data and estimate the ages.  Specifically, we question whether the adopted assumption of scaled-solar helium abundance is a valid approximation for the chemical evolution of the Galactic bulge at the metal-rich end. We demonstrate that the resulting impact on stellar parameter determination leads to a bias in inferred ages: photometric determinations are too old and spectroscopic determinations are too young.

\begin{figure}[H]
\begin{center}
\includegraphics[totalheight=0.58\textheight]{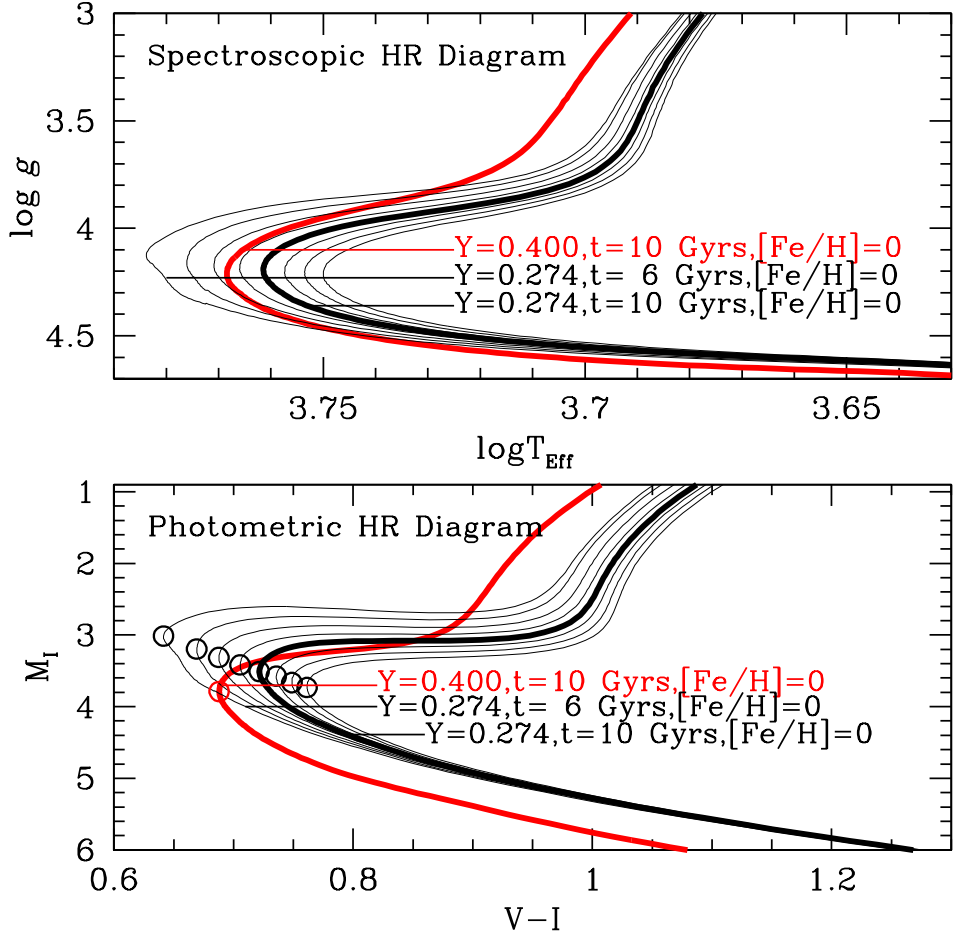}
\end{center}
\caption{Applying standard isochrones to helium-enhanced populations overestimates photometric ages and underestimates spectroscopic ages. TOP: [Fe/H]$=$0, Y$=$0.274 (solar) isochrones with ages 6-13 Gyr in black (thicker for t$=$10 Gyr), and an [Fe/H]$=$0, Y$=$0.400 (enhanced), t$=$10 Gyr isochrone plotted in red. On the spectroscopic plan, helium-enhanced populations mimic younger, standard populations with the effect minimized near the MSTO. BOTTOM: Same color scheme as above, but with tracks shown on the photometric $(V-I,I)$ plane. The 10 Gyr helium-enhanced track has a much fainter MSTO, equivalent to a t$\sim$13.5 Gyr track of the same metallicity and standard helium. A large difference in helium, ${\Delta}Y=$0.126,  has been selected for demonstrative purposes of clarity. Isochrones are from the Dartmouth stellar evolution database \citep{2008ApJS..178...89D}. } 
\label{Fig:Dartmouth4}
\end{figure}

\section{The Degeneracy Between Age and Helium}
\label{sec:HowHelium}
In general, isochrones assume the following prescription for initial helium:
\begin{equation}
Y = Y_{\rm{P}} + {\biggl[\frac{{\Delta}Y}{{\Delta}Z}\biggl]}_{\odot}Z,
\label{EQ:Iisochrone}
\end{equation}
\begin{equation}
Z = \frac{1 - Y_{\rm{P}}}{1 + {\biggl[\frac{{\Delta}Y}{{\Delta}Z}\biggl]}_{\odot}   + \frac{1}{(Z/X)}_{\odot}\times 10^{-\rm{[Fe/H]}}} \sim Z_{\odot}\times 10^{\rm{[Fe/H]}},
\end{equation}
where $Y$ is the initial helium abundance of the star, $Y_{\rm{P}}$ is the primordial helium abundance from big bang nucleosynthesis, $Z$ is the metallicity mass fraction of the star, and $({\Delta}Y/{\Delta}Z)_{\odot}$ is the slope derived by fitting a line from the primordial values, $(Z_{\rm{P}}, Y_{\rm{P}}) = (0,0.249)$ \citep{2008JCAP...08..011S},  to the solar values, $(Z_{\odot}, Y_{\odot}) \approx (0.018,0.272)$ \citep{2011arXiv1108.2273V}, yielding  $({\Delta}Y/{\Delta}Z)_{\odot} \approx1.5$. Some variation occurs as some isochrones use the pre-WMAP value  $Y_{P}=0.235$. Equation (\ref{EQ:Iisochrone}) is manifestly problematic. There is no \textit{a priori} reason that $Y$ should be a single-parameter, deterministic function of $Z$; that such a function should be a first-order polynomial; or that ${\Delta}Y/{\Delta}Z$ should be a universal constant. Why is the most abundant non-trivial element not generally incorporated as a variable input into isochrones? It is simply too difficult to measure: as the noble gas with two protons, it has the highest first ionization potential of any element, 24.6 eV. For the purposes of this work, we define a helium-enhanced stellar population as being a stellar population with values of $Y$ exceeding those predicted by Equation (\ref{EQ:Iisochrone}).

The effect of helium abundance on stellar evolution and thus inferred ages can be significant, due to the higher mean molecular weight and lower initial central hydrogen abundance accelerating the evolution of helium-enhanced stars. Hence, applying standard isochrones to helium-enhanced populations leads one to overestimate photometric ages \citep{2010ApJ...714.1072M}. And further, as we will demonstrate, the opposite holds for spectroscopic determinations: ages will be underestimated. 
 At fixed age and metallicity, MSTO stars with higher initial helium will be hotter and more compact, and thus bluer and dimmer. Their higher temperatures mimic younger ages on a $\log{g}$-$T_{\rm{eff}}$ diagram, but their smaller sizes and thus lower luminosity yield a dimmer MSTO, mimicking older ages on a CMD. On the SGB, the temperatures and bolometric luminosities become more similar, but not the surface gravities, causing spectroscopic ages to be underestimated. These effects are easily discerned in Figure \ref{Fig:Dartmouth4}. 

The assumption that ${\Delta}$Y/${\Delta}$Z is a constant has been demonstrated to be a spectacular failure in GCs. For example, $\omega$ Cen, NGC 2808, and 47 Tuc have a strong diversity of morphology in their CMDs, a fact that is well-explained by invoking helium-enriched second stellar generations \citep{2005ApJ...621..777P,2007ApJ...661L..53P,2009A&A...505..117C,2010MNRAS.408..999D,2011ApJ...736...94N,2011arXiv1109.2118N,2011arXiv1109.0900M,2011A&A...533A.120V}. Though there is no obvious causal implication for the bulge, GCs do empirically demonstrate that ${\Delta}Y/{\Delta}$Z can indeed vary. 

\section{The Case for Elevated Helium Enrichment in the Bulge}
\label{sec:CaseHelium}
The helium abundance of the bulge had been investigated in prior decades by comparing the number of red giant (RG) branch stars to HB stars. \citet{1988AJ.....96..884T}, \citet{1994A&A...285L...5R}, and \citet{1995A&A...300..109M} respectively estimated $Y=0.30\pm0.05$, $Y=0.30-0.35$, and $Y=0.28\pm0.02$. More recently, \citet{2011ApJ...730..118N,2011arXiv1109.2118N} found evidence for a helium-enhanced bulge using a different diagnostic: the red giant branch bump (RGBB). The bulge RGBB was shown to have anomalously low number counts $({\Delta}\ln(n) = -0.224 \pm 0.080)$ and high luminosity  $({\Delta}I = -0.104\pm0.045)$ relative to expectations from Galactic GCs and stellar models, consistent with an enrichment ${\Delta}Y\approx+0.06$ at the median metallicity. Both these methods may require incorporation of new evidence on RG evolution. \citet{2011ApJ...730...67B} demonstrated that 20-30\% of low-mass white dwarfs are single, implying that a significant number of stars might skip the HB phase, suggesting the above work may have systematically underestimated the bulge helium abundance.

The bulge should be expected to have a higher ${\Delta}$Y/${\Delta}$Z than the Sun due to its elevated $\alpha$-abundances \citep{2007arXiv0708.2445C}. \citet{2007ApJ...661.1152F},  \citet{2010A&A...513A..35A}, \citet{2011A&A...533A.134B}, \citet{2011A&A...530A..54G}, \citet{2011ApJ...732..108J} and \citet{2011arXiv1112.0306R} each find $\alpha$-abundance patterns elevated relative to the trends of the thin-disk, and possibly the thick-disk. 
The thin disk is the only stellar population for which the helium-metallicity relation has precise measurements (e.g., the Sun).  Elevated [$\alpha$/Fe] implies a lower relative contribution to chemical enrichment from type Ia SNe.  Since these explosions have an effective ${\Delta}Y/{\Delta}$Z$\approx0$, reducing their relative contribution increases ${\Delta}Y/{\Delta}$Z. Additionally, a low contribution from type Ia SNe implies a rapid timescale for star formation \citep{2010ApJ...723..329H}. This in turn implies a higher relative contribution from the asymptotic giant branch (AGB) ejecta of higher mass stars, which are more helium-enriched than those of low and intermediate mass stars that take longer to reach the AGB. \cite{2008MNRAS.391..354R} showed that the abundance of gas ejected from 3-10 $M_{\odot}$ AGB stars is $\langle Y\rangle=0.33$, with the total yield and  composition being a function of the metallicity and the slope of the initial mass function. Further, stellar models predict that the winds of fast rotating massive stars (WFRMS) are extremely helium-enriched,  with values reaching $\langle Y\rangle=0.50$ \citep{2007A&A...464.1029D}. Thus, the $\alpha$-enhancement of the bulge implies a higher effective value of ${\Delta}Y/{\Delta}Z$ due to two distinct causes. As the duration of star formation increases, lower-mass AGB ejecta, which are less He-enriched, will contribute relatively more to the chemical evolution budget than those of higher mass AGB+WFRMS ejecta. Second, late-time Type Ia SNe, explosions that contribute a lot of metals but no helium, will begin to go off. 
Figure 1 of \citet{2007arXiv0708.2445C} implies Y$\approx$0.31 at [Fe/H]$=$0 for a population with an $\alpha$-enhancement similar to the bulge. This is  midway between standard isochrones values and the estimate of \citet{2011arXiv1109.2118N}. We note that there has been some skepticism regarding WFRMS as the source of GC helium because they enrich metals as well as helium, and because GCs have a weak gravitational potential well that may not hold on to these ejecta. Neither of those concerns apply to the bulge.  The abundances from \citet{2011A&A...533A.134B} suggest that the sodium-oxygen anti-correlation seen in GCs and thought to trace high-mass AGB and/or WFRMS ejecta, is present in metal-rich bulge stars and trends \textit{with} [Fe/H]. 

We refer to a new class of models \citep{2011ApJ...740L..45C,2012ApJ...747...78B} motivated by the need to explain the UV-upturn of old bulges and ellipticals, which are best explained by the presence of blue horizontal branch stars. \citet{2011ApJ...740L..45C} and \citet{2012ApJ...747...78B} both posit imperfect mixing between AGB and SNe ejecta, thereby having stars form directly from the gas of helium-rich AGB ejecta without dilution from SNe ejecta. If this process occured in other spheroids, it likely also occured in the bulge, and vice versa. These models also make the testable prediction that there should be a large helium spread at fixed metallicity. 

There is another effect that may be at play: helium sedimentation. Simulations of hot gas in galaxy clusters by \citet{2004MNRAS.349L..13C} predict that helium preferentially accretes onto cluster cores, as:
\begin{equation}
\frac{{\Delta}Y}{Y} \approx \frac{4}{3}   \biggl(\frac{t}{10\;\rm{Gyr}}\biggl)         \biggl(\frac{f_{\rm{gas}}}{0.1}\biggl)^{-1}      \biggl(\frac{T}{10\;\rm{KeV}}\biggl)^{3/2}.
\end{equation}
They argue that values of ${\Delta}Y\approx0.12$  can be reached by cD galaxies, which could explain the UV upturn of these galaxies by creating more extreme blue HB stars. While the effect in the bulge should be much smaller, it still may contribute to helium enhancement.


\newpage
\section{Evidence of Enhanced Helium in the Analysis of Spectroscopic Data and Predictions Thereof}
\label{sec:Systematic}
\begin{figure}[H]
\begin{center}
\includegraphics[totalheight=0.7\textheight]{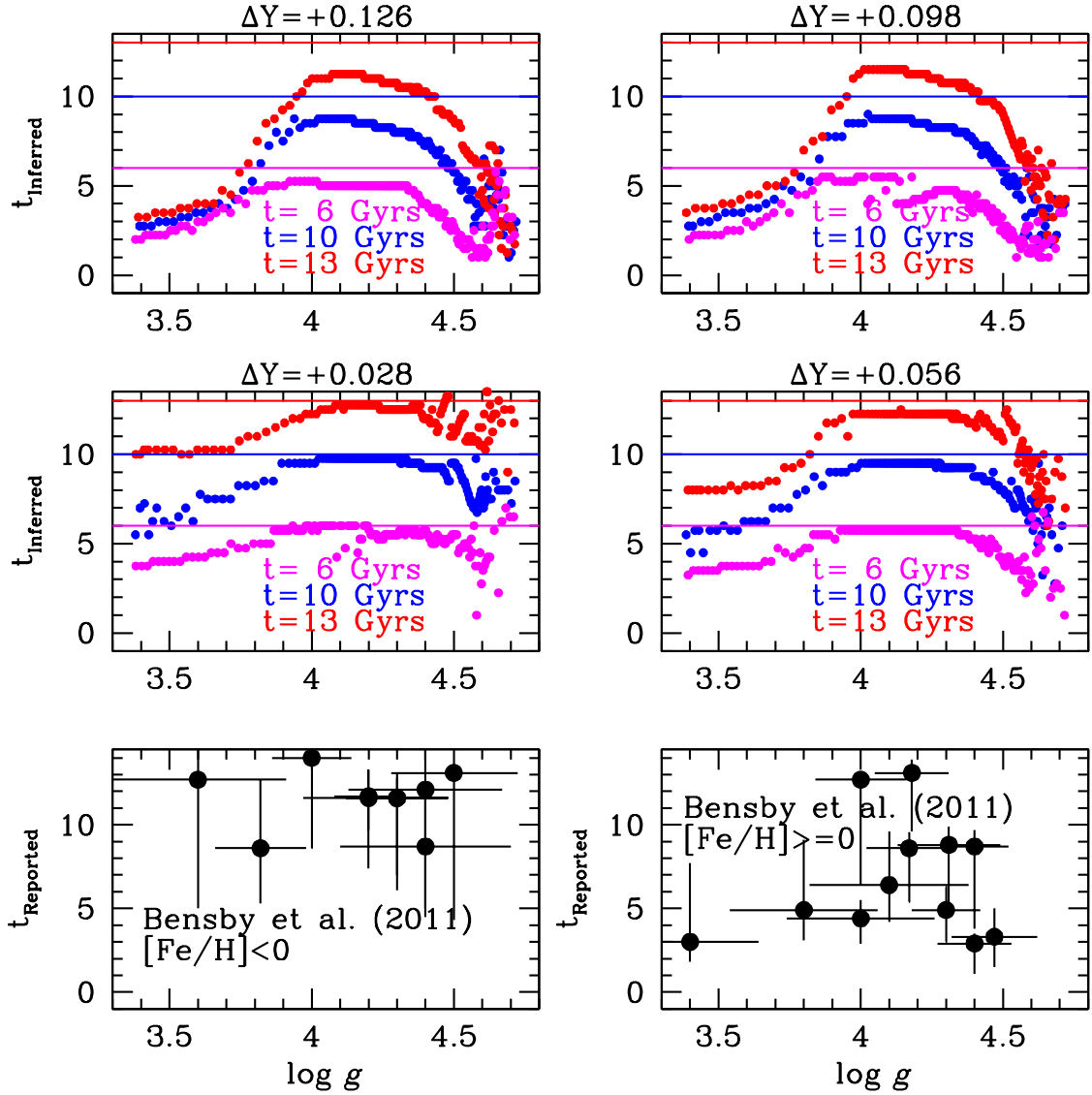}
\end{center}
\caption{Applying standard isochrones to helium-enhanced populations leads to an offset between the inferred and true age that is monotonic in ${\Delta}Y$ and is minimized near the MSTO. Top four panels show the inferred age for stars 6 Gyr old (magenta), 10 Gyr old (blue) and 13 Gyr  (red) by standard isochrones as a function of $\log{g}$, for different values of helium-enhancement. The true ages are shown by the horizontal lines to guide the eye. Bottom panels shows the reported ages as a function of $\log{g}$ from \citet{2011A&A...533A.134B}, separated into subsolar and supersolar samples. The metal-poor stars have a flat $\log{g}$-age distribution, whereas the metal-rich stars show older ages near the MSTO.} 
\label{Fig:logg}
\end{figure}

\begin{figure}[H]
\begin{center}
\includegraphics[totalheight=0.7\textheight]{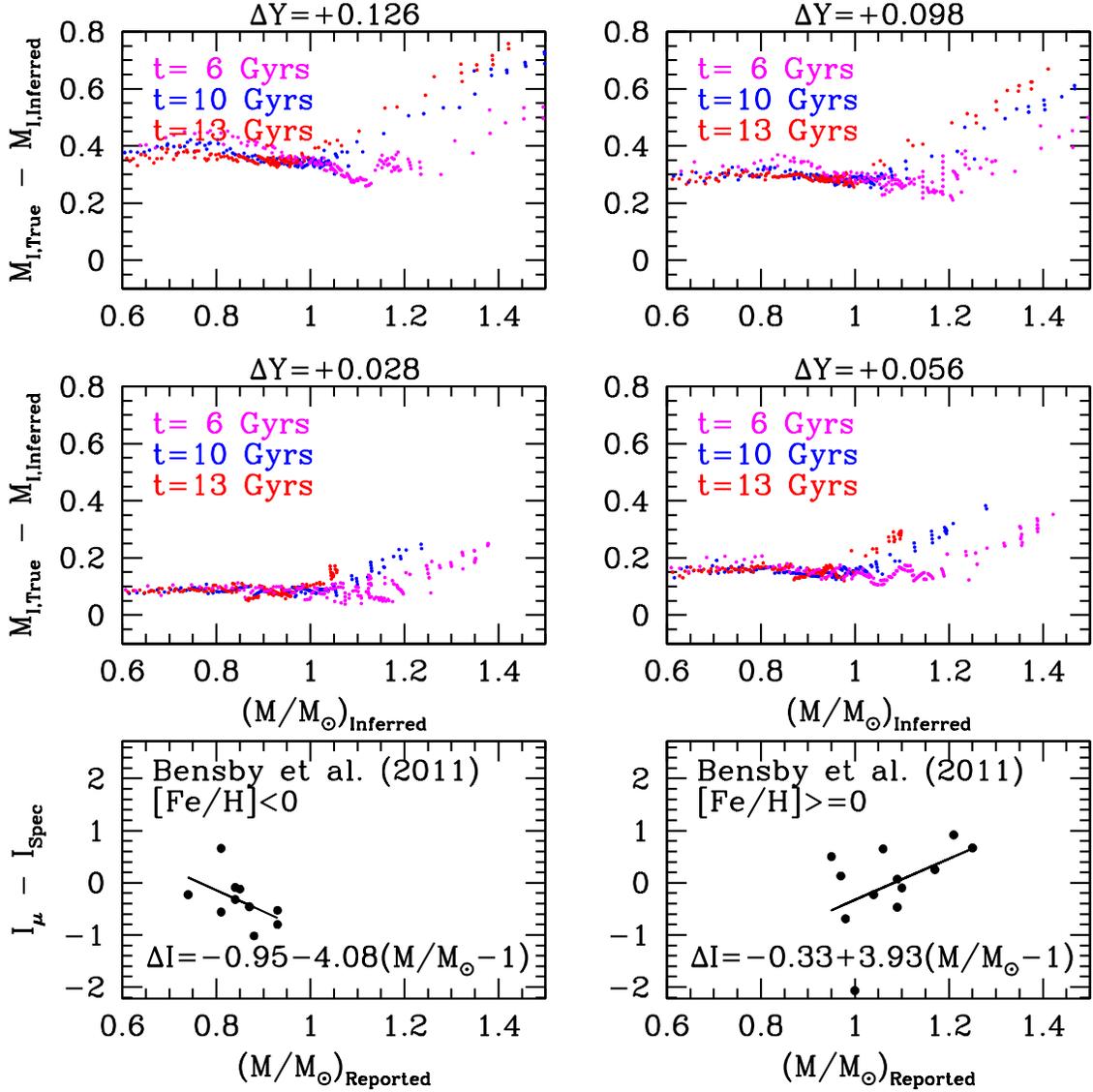}
\end{center}
\caption{Top four panels show the predicted error in the absolute magnitude inferred for stars 6 Gyr old (green), 10 Gyr old (blue) and 13 Gyr (red) by standard isochrones of the correct metallicity as a function of the inferred stellar mass, for different values of helium-enhancement. Bottom panels show the difference between absolute magnitude inferred from the microlensing lightcurves and the inferred absolute magnitude as a function of inferred spectroscopic mass for the sample of \citet{2011A&A...533A.134B}. The best-fit line to the metal-rich stars has both a greater intercept and a greater slope.} 
\label{Fig:Minferred}
\end{figure}

We discuss two systematic trends predicted by stellar models. First, we predict that the age offset induced by applying standard spectroscopic $(\log{g}-T_{\rm{eff}})$ isochrones should be minimized near the MSTO and maximized on the SGB: the highest spectroscopic ages should be for stars nearest to the MSTO.  Second, we predict that the error in the inferred absolute magnitude (M$_{I,\rm{true}}$ $-$ M$_{I,\rm{inferred}}$) should on average be positive, and correlate positively with the spectroscopically inferred mass. The predicted trends as well as the data of \citet{2011A&A...533A.134B} are shown in Figures \ref{Fig:logg} and \ref{Fig:Minferred}. Both predicted trends are consequences of the same underlying phenomenon; the size of the age offset being a function of evolutionary state. The theoretical predictions shown in Figure \ref{Fig:logg} are summarized in Table \ref{table:corrections}.

We use isochrones from the Dartmouth stellar evolution database \citep{2008ApJS..178...89D}. The 8 input stellar parameters are ($t=$10,13 Gyr, [Fe/H]$=$0, [$\alpha$/Fe]$=$0,$+$0.4, 0.33,0.40), where  ([Fe/H]$=$0, [$\alpha$/Fe]$=+$0.40) is approximately equivalent to  ([Fe/H]$=+$0.30, [$\alpha$/Fe]$=+$0.00) \citep{1993ApJ...414..580S}. The upper four panels in each figure show the effect of different offsets in helium: ${\Delta}Y=$0.028, 0.056, 0.098 and 0.126, for three assumed intrinsic ages, 6 Gyr (magenta), 10 Gyr (blue) and 13 Gyr (red). We evaluate these populations using standard isochrones ($Y=0.245 + 1.6$Z) from the same database. When analyzing predicted stellar observables of one age and composition with isochrones of a different age and composition, we use the metric:
\begin{equation}
{\Delta}S^2 = {\biggl[\frac{{\Delta}T}{92\rm{\;K}}\biggl]}^2 + {\biggl[\frac{{\Delta}\log{g}}{0.195\rm{\;dex}}\biggl]}^2,
\end{equation}
where the normalizations are the average errors reported by \citet{2011A&A...533A.134B}. Inspection of Figures \ref{Fig:logg} and \ref{Fig:Minferred} reveals that the predicted effects are monotonic with ${\Delta}Y$ at fixed $\log{g}$ or $(M/{M_{\odot}})_{\rm{Inferred}}$. 

\newpage
\begin{table*}[ht]
\caption{Predicted values of $t_{\rm{Inferred}}$/$t_{\rm{True}}$ as a function of $\log(g)$ and ${\Delta}Y$. The first four columns are for the $t_{\rm{True}}=10$ Gyr case, and the last four columns are for the  $t_{\rm{True}}=13$ Gyr case. \newline}
\begin{center}
\scalebox{0.95}
{\begin{tabular}{| l | l l l  l | l l l l |}
\hline\hline\hline
 $\log(g)$ $\backslash$ ${\Delta}Y$ & 0.028 & 0.056 & 0.098 & 0.126 & 0.028 & 0.056 & 0.098 & 0.126 \\
\hline\hline
3.440000 & 0.64 & 0.54 & 0.30 & 0.29 & 0.78 & 0.62 & 0.29 & 0.26 \\ 
3.480000 & 0.66 & 0.56 & 0.31 & 0.30 & 0.78 & 0.62 & 0.30 & 0.27 \\ 
3.520000 & 0.66 & 0.58 & 0.33 & 0.32 & 0.78 & 0.62 & 0.31 & 0.28 \\ 
3.560000 & 0.69 & 0.59 & 0.35 & 0.33 & 0.78 & 0.63 & 0.33 & 0.29 \\ 
3.600000 & 0.70 & 0.62 & 0.38 & 0.35 & 0.79 & 0.64 & 0.34 & 0.32 \\ 
3.640000 & 0.73 & 0.65 & 0.41 & 0.38 & 0.80 & 0.65 & 0.37 & 0.34 \\ 
3.680000 & 0.76 & 0.68 & 0.43 & 0.42 & 0.80 & 0.67 & 0.40 & 0.39 \\ 
3.720000 & 0.78 & 0.71 & 0.48 & 0.47 & 0.82 & 0.70 & 0.44 & 0.44 \\ 
3.760000 & 0.81 & 0.75 & 0.53 & 0.53 & 0.83 & 0.74 & 0.49 & 0.49 \\ 
3.800000 & 0.84 & 0.79 & 0.58 & 0.59 & 0.85 & 0.78 & 0.54 & 0.54 \\ 
3.840000 & 0.87 & 0.83 & 0.64 & 0.65 & 0.87 & 0.81 & 0.60 & 0.60 \\ 
3.880000 & 0.90 & 0.86 & 0.70 & 0.72 & 0.89 & 0.85 & 0.66 & 0.67 \\ 
3.920000 & 0.92 & 0.89 & 0.76 & 0.78 & 0.91 & 0.88 & 0.73 & 0.73 \\ 
3.960000 & 0.94 & 0.91 & 0.81 & 0.83 & 0.93 & 0.91 & 0.79 & 0.78 \\ 
4.000000 & 0.96 & 0.93 & 0.85 & 0.85 & 0.94 & 0.93 & 0.84 & 0.81 \\ 
4.040000 & 0.97 & 0.95 & 0.87 & 0.87 & 0.96 & 0.94 & 0.87 & 0.84 \\ 
4.080000 & 0.97 & 0.95 & 0.88 & 0.88 & 0.97 & 0.94 & 0.88 & 0.86 \\ 
4.120000 & 0.98 & 0.95 & 0.87 & 0.87 & 0.98 & 0.94 & 0.88 & 0.86 \\ 
4.160000 & 0.97 & 0.95 & 0.87 & 0.86 & 0.98 & 0.94 & 0.88 & 0.86 \\ 
4.200000 & 0.97 & 0.95 & 0.86 & 0.84 & 0.98 & 0.94 & 0.87 & 0.85 \\ 
4.240000 & 0.97 & 0.95 & 0.84 & 0.83 & 0.97 & 0.94 & 0.85 & 0.84 \\ 
4.280000 & 0.97 & 0.94 & 0.82 & 0.81 & 0.96 & 0.94 & 0.84 & 0.82 \\ 
4.320000 & 0.96 & 0.93 & 0.82 & 0.80 & 0.96 & 0.94 & 0.83 & 0.81 \\ 
4.360000 & 0.95 & 0.92 & 0.79 & 0.77 & 0.96 & 0.92 & 0.82 & 0.80 \\ 
4.400000 & 0.93 & 0.89 & 0.74 & 0.74 & 0.94 & 0.91 & 0.78 & 0.78 \\ 
4.440000 & 0.92 & 0.86 & 0.69 & 0.69 & 0.93 & 0.90 & 0.75 & 0.73 \\ 
4.480000 & 0.91 & 0.82 & 0.63 & 0.62 & 0.93 & 0.88 & 0.72 & 0.68 \\ 
4.520000 & 0.89 & 0.77 & 0.54 & 0.53 & 0.89 & 0.89 & 0.63 & 0.61 \\ 
4.560000 & 0.77 & 0.73 & 0.44 & 0.42 & 0.84 & 0.79 & 0.52 & 0.50 \\ 
4.600000 & 0.76 & 0.67 & 0.34 & 0.41 & 0.88 & 0.72 & 0.44 & 0.42 \\ 
\hline
\end{tabular}
\label{table:corrections}
}\end{center}
\end{table*}

\section{Preliminary constraints of the Age and Helium Abundance of the Bulge}
\label{sec:Constraints}
We first estimate ${\Delta}Y$ given the assumption that isochrones with the ``correct'' helium values would yield ages on the MSTO and SGB that are independent of $\log{g}$. We test for this by rescaling the ages (and their errors) using the values of $t_{\rm{inferred}}/t_{\rm{true}}$ from Table \ref{table:corrections}, and we compute ${\chi}^2$:
\begin{equation}
{\chi}^2 = \sum_{i} \biggl(\frac{\overline{Age}-Age(i))}{\sigma_{Age(i)}}\biggl)^2,
\end{equation}
We obtain ${\chi}^2 =$ 30.3, 28.2, 27.2, 24.8 and 25.6 for ${\Delta}Y =$ 0.0, 0.028, 0.056, 0.098, and 0.126. Thus ${\chi}^2$ is minimized for ${\Delta}Y = 0.098$, yielding ${\Delta}{\chi}^2 = 5.4$ -- a 2.3$\sigma$ detection of helium-enhancement. The inferred enhancement is therefore ${\Delta}Y=0.098 \pm 0.043$ at [Fe/H] $\approx +$0.3. In contrast, applying the same procedure to metal-poor stars \textit{increases }${\chi}^2$.

We now estimate, using a different method, an \textit{upper-bound} on ${\Delta}Y$. We first impose the requirement that the photometric and spectroscopic ages agree. From Figure \ref{Fig:Dartmouth4}, we see that a 10 Gyr isochrone with ${\Delta}Y=+0.126$ has an MSTO as faint as that of a standard $\sim$13.5 Gyr isochrone, for a 35\% offset, thus the photometric age error can be linearly approximated by:
\begin{equation}
\biggl(\frac{t_{\rm{Inferred}}}{t_{\rm{True}}}\biggl)_{Phot} \approx \biggl(1+2.8{\Delta}Y\biggl),
\end{equation}
For the spectroscopic age inference, the age offset is a function of evolutionary state, so we estimate by taking the mean value of $t_{\rm{Inferred}}/t_{\rm{True}}$ in the range $3.44 \leq \log{g} \leq 4.60$:
\begin{equation}
\biggl\langle  \frac{t_{\rm{inferred}}}{t_{\rm{True}}}\biggl\rangle_{Spec} \approx \biggl(1-2.9{\Delta}Y\biggl),
\end{equation}
and from the literature \citep{2011A&A...533A.134B,2011ApJ...735...37C}, we have that:
\begin{equation}
\frac{t_{\rm{Photometric}}}{t_{\rm{Spectroscopic}}}  \lesssim 2.0 \approx \frac{1+2.8{\Delta}Y}{1-2.9{\Delta}Y}
\end{equation}
Combining these equations yields a helium enrichment relative to isochrones of ${\Delta}Y \lesssim 0.11$ for the metal-rich stars, yielding a corresponding mean age of $t_{\rm{Bulge}} \approx 10$ Gyr, and a value of $({\Delta}Y/{\Delta}Z)_{\rm{Bulge}} \lesssim 5.0$ if one assumes a linear form for $Y(Z)$. As the canonical helium abundance for stars at [Fe/H]$\approx+0.30$ is $Y\approx0.31$, this sets an upper bound of $Y\approx0.42$. This upper bound on ${\Delta}Y$ is derived by equating the ages of the metal-rich bulge stars with the mean age of the bulge: it assumes the \textit{limiting condition} of instantaneous star formation. Any age spread will decrease the amount of extra helium required. In contrast, \citet{2011A&A...533A.134B} derive an age difference between the metal-poor and metal-rich stars of $\sim$6 Gyr, under the assumption of standard helium enrichment.




\section{Discussion}
\label{sec:Discussion}

We have shown in this \textit{Letter} that the discrepancy in Galactic bulge turnoff age estimates can be well-explained by the hypothesis that the chemical evolution of the Galactic bulge is helium-enhanced relative to that assumed by standard isochrones. We first found evidence for helium-enhancement at the 2.3$\sigma$ level, suggesting ${\Delta}Y=+0.098$ for the metal-rich stars. We also derived an upper bound on the helium enrichment of the bulge, ${\Delta}Y \lesssim 0.11$ for the metal-rich ([Fe/H]$\approx+0.30$) stars. We deliver two testable predictions, that the absolute magnitudes will be systematically fainter than the best-fit values from standard isochrones, and that inferred ages will be highest near the MSTO. Both diagnostics suggest helium-enrichment for the metal-rich stars, with no such evidence for the metal-poor stars.

There are multiple roads forward. On the theory side, chemical evolution models will be required to ascertain what conditions are required in the primeval Milky Way for a high value of $({\Delta}Y/{\Delta}Z)_{\rm{Bulge}}$. A broader range of isochrones than currently available will be required to interpret the available photometric and spectroscopic data.  Finally, it is possible that the stellar atmosphere models used by spectroscopists fail in the case of atmospheres that are extremely helium-enriched.  The [Fe/H] distribution of MSTO+SGB stars \citep{2011A&A...533A.134B} is inconsistent with that derived from giants \citep{2008A&A...486..177Z,2011A&A...534A..80H,2011ApJ...732..108J}, as it has a trough near [Fe/H]$=$0, where the giant metallicities show a peak. Could this be the reason? 
Observationally, an uncertainty in the value of $({\Delta}Y/{\Delta}Z)_{\rm{Bulge}}$ should be incorporated in analyses of bulge stars. The age discrepancy discussed in this paper constitutes the fourth, independent line of evidence for a helium-enriched bulge, following previous investigations of star counts on the upper RG branch, on the RGBB, and chemical evolution arguments based on the high values of [$\alpha$/Fe].

The helium abundances suggested here are difficult to attain with chemical evolution models, as ${\Delta}Y{\Delta}Z \lesssim 3.0$ for general assumptions \citep{2002A&A...395..789C,2008RMxAA..44..341C}. Should the inferred helium-enhancement for the bulge hold up under further scrutiny it may necessitate rethinking the underlying assumptions of chemical evolution models. Additionally, even in the disk, there may be surprises with respect to helium-enrichment: \citet{2010A&A...518A..13G} find ${\Delta}Y/{\Delta}Z=5.3\pm1.4$ in a sample of local K-dwarfs. 

The preliminary estimates of this Letter result from the assumption of instantaneous star formation. How much age spread is reasonable depends on assumptions of the initial conditions of the Galaxy. The assumption of rapid gravitational infall yields ${\Delta}t_{\rm{Bulge}}\approx0.5$ Gyr \citep{2011A&A...525A.126C}, whereas the ``clump-origin bulge'' model, in which the bulge is formed by stellar clumps forming in a gas-rich disk and accreting to its center by dynamical friction, yields ${\Delta}t_{\rm{Bulge}}\approx2.0$ Gyr \citep{2011arXiv1109.2898I}. However, recent investigations of star counts and radial velocities find that bulge kinematics are consistent with a pure N-body bar that evolved from secular instabilities, which would imply that bulge stars are just disk stars on bar orbits \citep{2010ApJ...720L..72S,2011ApJ...734L..20M}. This would suggest a duration of star formation as extended as that of the inner disk.


The results of these endeavors may have significant cosmological implications. Population synthesis studies of other galaxies assume a solar value of ${\Delta}Y/{\Delta}$Z. If this parameter is demonstrated to be non-universal for field stars, it will affect age and mass determinations of field galaxies.

\acknowledgments
DMN was primarily supported by the NSERC grant PGSD3-403304-2011. DMN and AG were partially supported by the NSF grant AST-1103471. We thank Jennifer A. Johnson for enlightening discussions.

\end{document}